\documentclass[12pt,preprint]{aastex}

\shorttitle{Gaseous CO$_2$ toward Cepheus A East}
\shortauthors{Sonnentrucker et al.}

\begin{document}


\title{Gas-phase CO$_2$ emission toward Cepheus A East: \\
the result of shock activity?\altaffilmark{1}}


\author{P. Sonnentrucker\altaffilmark{2}, E. Gonz\'alez-Alfonso\altaffilmark{3}, D. A. Neufeld\altaffilmark{2}, 
E. A. Bergin\altaffilmark{4}, G. J. Melnick\altaffilmark{5}, \\
W. J. Forrest\altaffilmark{6}, J. L. Pipher\altaffilmark{6}, and D. M. Watson\altaffilmark{6}}


\altaffiltext{1}{Based on observations with the {\it Spitzer Space Telescope}.}
\altaffiltext{2}{Department of Physics and Astronomy, Johns Hopkins University, 3400 North Charles Street, Baltimore, MD 21218.}
\altaffiltext{3}{Universidad de Alcal\'a de Henares, Departamento de F\'isica, Campus Universitario, E-28871
 Alcal\'a de Henares, Madrid, Spain.}
\altaffiltext{4}{Department of Astronomy, University of Michigan, 825 Dennison Building, 
500 Church Street, Ann Arbor, MI 48109.}
\altaffiltext{5}{Harvard-Smithsonian Center for Astrophysics, 60 Garden Street, Cambridge, MA 02138.}
\altaffiltext{6}{Department of Physics and Astronomy, University of Rochester, Rochester, NY 14627.}


\begin{abstract}

We report the first detection of gas-phase CO$_2$ emission in the star-forming region Cepheus A East, obtained by spectral line mapping of the $\nu_2$ bending mode at 14.98 $\mu$m with the Infrared Spectrograph (IRS) instrument onboard the {\it Spitzer Space Telescope}. The gaseous CO$_2$ emission covers a region about 35$''$ $\times$ 25$''$ in extent, and results from radiative pumping by 15 $\mu$m continuum photons emanating predominantly from the HW2 protostellar region. The gaseous CO$_2$ exhibits a temperature distribution ranging from 50 K to 200 K. A correlation between the gas-phase CO$_2$ distribution and that of H$_2$ S(2), a tracer of shock activity, indicates that the CO$_2$ molecules originate in a cool post-shock gas component associated with the outflow powered by HW2. The presence of CO$_2$ ice absorption features at 15.20 $\mu$m toward this region and the lack of correlation between the IR continuum emission and the CO$_2$ gas emission distribution further suggest that the gaseous CO$_2$ molecules are mainly sputtered off grain mantles -- by the passage of slow non-dissociative shocks with velocities of 15-30 km s$^{-1}$ -- rather than sublimated through grain heating. 

\end{abstract}


\keywords{ISM: Molecules --- ISM: Clouds, molecular processes --- Star-forming region: individual (Cepheus A East)}



\section{Introduction}

Cepheus A is a well-known site of star formation that has been studied extensively. Previous results have shown that it contains a series of heavily embedded far-infrared and radio-continuum sources, one of which dominates the luminosity of the entire region (HW2 with 2.5$\times$10$^{4}$ $L_{\odot}$; Hughes \& Wouterloot 1984; Evans et al. 1981). Ground- (e.g. Goetz et al. 1998) and space-based observations (e.g. Hartigan et al. 1996; Wright et al. 1996; van den Ancker et al. 2000) revealed that the multipolar outflow existing in this region consists of complex structures of shock-excited atomic and molecular gas. H$_2$ emissions and line emissions from ionized species were detected along high-velocity (HV) and extra-high-velocity (EHV) outflows, indicating the presence of both dissociative (J-type) and non-dissociative (C-type) shocks, likely resulting from successive episodes of activity (Narayanan \& Walker 1996). The protostellar object HW2 was determined to be the dominant powering source of the EHV outflow, while the source of the HV jet is still under debate (Goetz et al. 1998; Hiriart et al. 2004). The cavities carved into the surrounding molecular gas by the quadrupolar outflow are clearly seen in the NH$_3$ maps obtained by Torrelles et al. (1993).

In this {\it Letter}, we focus on new observations of Cepheus A East obtained with the Infrared Spectrograph (IRS) onboard the {\it Spitzer Space Telescope} ({\it SST}).  The benefit of the {\it SST} data lies in the much greater sensitivity and the significantly better spatial resolution offered by the IRS compared to the {\it Infrared Space Observatory (ISO)}. Such advances allow for the detection and spectral mapping of variations in the molecular and atomic line intensities on scales of a few arcsec instead of the previous few tens of arcsec or few arcmin scales, thus enabling our understanding of shock physics and chemistry at much finer scales. Detections of H$_2$ S(1) through S(7), C$_2$H$_2$, [\ion{Ne}{2}], [\ion{Ne}{3}], [\ion{S}{1}], [\ion{S}{3}] and [\ion{Fe}{2}] emissions as well as absorption from CO$_2$ and H$_2$O ices will be discussed in a future paper. The present work will focus on the first detection of gas-phase CO$_2$ emission toward Cepheus A East. Section 2 describes the observations and data analysis. Sections 3 \& 4 discuss our new results as well as the spatial distributions of gas-phase CO$_2$ and H$_2$ S(2) in the context of shock chemistry and outflow activity. 

\section{Observations and data reduction}


Spectral maps of two overlapping 1$'$ $\times$ 1$'$-square fields were obtained toward Cepheus A East with the IRS instrument onboard the {\it Spitzer Space Telescope (SST)} as part of Guaranteed Time Observer (GTO) program 113. The Short-Low (SL, both orders), Short-High (SH) and Long-High (LH) modules allowed for wavelength coverage from 5.2 to 25 $\mu$m. Most of the data longward of 25 $\mu$m suffer severe detector saturation as a result of very strong continuum emission and were not used. Continuous spatial coverage in the overlapping fields was obtained by stepping the slit perpendicular and parrallel to its long axis in steps of one-half its width and 4/5 its length, respectively. 

The data were processed at the Spitzer Science Center (SSC) using version 12 of the IRS pipeline. The Spectroscopy Modeling Analysis and Reduction Tool (SMART) software (Higdon et al. 2004) was used to extract wavelength-calibrated spectra at each spatial position in the maps. We used a set of locally-developed IDL routines specifically designed to remove bad pixels in the high resolution modules, to calibrate the fluxes for extended sources, and to generate spectral line maps from the extracted spectra (Neufeld et al. 2006).




\section{Results}

Figure~1 shows two examples of spectral line intensity maps we obtained using the SH data alone to allow direct comparison of the H$_2$ S(2) and gaseous CO$_2$ spatial distributions. The right panel displays the spatial distribution of the H$_2$ S(2) intensity (12.28 $\mu$m), while the left panel shows the intensity distribution for gas-phase CO$_2$ ($Q$-branch of the $\nu_2$ bending mode at 14.98 $\mu$m). The H$_2$ and the CO$_2$ emissions mainly arise at the surfaces of the NH$_3$-free cavities carved by the outflows. These emissions thus seem to result from interactions between the EHV outflow and the quiescent molecular gas traced by NH$_3$ (G\'omez et al. 1999). Figure~2 (left panel) displays examples of summed spectra in the wavelength range relevant for this study. 

To constrain the physical conditions in the CO$_2$-containing gas, we generated synthetic profiles of the $\nu_2$ CO$_2$ band for temperatures ranging from 50 K to 900 K and compared them with the observed spectra (Fig.~2, right panel). We find that temperatures between 50 and 200 K best fit the observed CO$_2$ $Q$-branch emission (14.98 $\mu$m) over the entire CO$_2$-emitting region. Our observations, therefore, reveal the presence of a post-shock gas component much cooler than that measured with H$_2$ pure rotational line transitions (e.g. van den Ancker et al. 2000, $T \sim$ 730 K from H$_2$ S(1)--S(5) over the {\it ISO} Short Wavelength Spectrometer beam sizes 14$''$$\times$20$''$ and 14$''$$\times$27$''$). 

To determine the column density associated with the gas-phase CO$_2$ emission, one needs to identify the excitation mechanism. Prior studies (e.g., Gonz\'alez-Alfonso \& Cernicharo 1999; Boonman et al. 2003) showed that gaseous CO$_2$ molecules can be excited: 1) by collisions with H and H$_2$; 2) by radiative pumping into the 4.27 $\mu$m band; 3) by radiative pumping due to 15 $\mu$m continuum photons emitted by dust local either to the CO$_2$ gas component or to HW2. 

For collisions to dominate, densities in excess of $n \sim$ 10$^{8}$ cm$^{-3}$ would be required. However, prior observations indicate densities from few$\sim$10$^3$ to few$\times$10$^7$ cm$^{-3}$ in the outflow region (Codella et al. 2005). Radiative pumping into the 4.27 $\mu$m band and subsequent cascade can be disregarded because the relaxation to the 15 $\mu$m band would produce detectable emission at 13.9 $\mu$m and 16.2 $\mu$m that we do not observe. Resonant scattering by 15 $\mu$m continuum photons is, therefore, the most likely mechanism to excite the observed gaseous CO$_2$ molecules. Continuum maps obtained from our data also indicate that the radiation field is dominated by dust emission close to HW2 -- from a reflection nebula (Casement \& McLean 1996; Goetz et al. 1998, Mart\'in-Pintado et al. 2005)-- rather than by dust emission local to the CO$_2$ gas component. We, hence, conclude that the gas-phase CO$_2$ molecules detected toward Cepheus A East are predominantly excited via radiative pumping by 15 $\mu$m continuum photons emanating from the protostellar region HW2. Resonant scattering in the CO and H$_2$O vibrational bands, through continuum photons emanating from IRc2/BN, was also found toward the shock region Orion Peak 1/2 (Gonz\'alez-Alfonso et al. 2002).

To compute $N$(CO$_2$), we adopted the total luminosity and source effective temperature from Lenzen et al. (1984), as well as a distance of 690 pc for the Cepheus A East region, and we assumed that the projected distance of each spatial position is equal to its true distance from HW2. Under that assumption, the inferred CO$_2$ column density is proportional to the measured intensity and to the square of the angular separation from the exciting source HW2. We also corrected the CO$_2$ intensity measurements for extinction caused by the CO$_2$ ice absorption. A comparison of the $N$(CO$_2$) map (Fig.~3, left panel) with the intensity distribution of the H$_2$ S(2) line ($I$(H$_2$) S(2), Fig.~1, right panel) indicates that -- like H$_2$ S(2) -- CO$_2$ peaks at the NE position, coincident with the NH$_3$ bridge between Cep-A2 and Cep-A3 (Torrelles et al. 1993) and at the HW5/HW6 positions. These striking similarities again strongly suggest that the CO$_2$ emission arises in post-shock gas, traced by the warm H$_2$ emission, and results from interactions between the EHV outflow and the ambient medium. The right panel of Fig.~3 displays the derived, extinction corrected, column density measurements of gas-phase CO$_2$ $versus$ H$_2$ S(2) for spatial positions where the CO$_2$ line intensity ($Q$-branch) is greater than 0.75$\times$10$^{-4}$ erg s$^{-1}$ cm$^{-2}$ sr$^{-1}$. Our estimate of the Spearman coefficient ($\rho \sim$ 0.42 and significance level $<$1\%) indicates that the column densities of the two species tend to increase together with spatial position.

\section{Discussion}

The presence of CO$_2$ ice absorption features at most spatial positions in the gaseous CO$_2$-emitting region, their progressive weakening along the EHV outflow axis, as well as the localization of the gaseous CO$_2$ emission into a cool post-shock component, all pose the question of the origin of the gas-phase CO$_2$ molecules in Cepheus A East. Was CO$_2$ injected into the gas-phase by sublimation from CO$_2$-rich dust mantles after substantial grain heating or by grain sputtering caused by the passage of slow non-dissociative shocks known to exist in this active region?

We searched for evidence of a correlation between the dust continuum emission, a tracer of grain heating, and the gaseous CO$_2$ distribution in the CO$_2$-emitting region by comparing the CO$_2$ column density measurements with measurements of the dust continuum emission at 7.3 $\mu$m, 14.8 $\mu$m and 18.0 $\mu$m. We found no obvious relation between $N$(CO$_2$) and the dust continuum emission at these wavelengths. Furthermore, we determined that the dust temperature is never greater than the CO$_2$ sublimation temperature ($\sim$ 90 K) at the locations of the CO$_2$ column peaks. While grain heating likely occurs in the active region close to HW2, CO$_2$-rich grain mantle sublimation is not the dominant source of gaseous CO$_2$ in Cepheus A East. Sputtering off CO$_2$-rich ice mantles is, therefore, the most likely scenario in the present case. Our measurements of the CO$_2$ ice column density range from $\sim$ 10$^{17}$ to $\sim$ few $\times$ 10$^{18}$ cm$^{-2}$ over the region exhibiting CO$_2$ gas emission, indicating that only a few percent of the material has passed through mantle-destroying shocks.

To derive the abundance of gaseous CO$_2$, an estimate of $N$(H$_2$) in the CO$_2$ gas component is needed. Because the gas-phase CO$_2$ molecules arise from a post-shock component much cooler than that containing the H$_2$ gas we detected, we cannot directly measure the total column density of the gas at $T=$ 50-200 K from our data. Our abundance estimate will, hence, rely on previous indirect measures. G\'omez et al. (1999) derived $N$(H$_2$) $=$ 1.5 $\times$ 10$^{22}$ cm$^{-2}$ in the outflow using HCO$^+$ emission measurements; a similar estimate was obtained by van den Ancker (2000) using far-infrared CO emission at the {\it ISO}/Long Wavelength Spectrometer angular resolution (75$''$ beam size). Adopting this H$_2$ column density yields an average gas-phase CO$_2$ abundance of few $\times$10$^{-7}$. 

This value should be regarded as a lower limit since the $N$(CO$_2$) might be underestimated if the true distance from HW2 is greater than the projected distance we assumed. A small fraction of the gaseous CO$_2$ might arise from CO$_2$ ice sublimation since some CO$_2$ ice profiles, especially close to HW2, show substructures characteristic of grain heating (Gerakines et al. 1999). A fraction of the gas-phase CO$_2$ emission might also come from quiescent gas associated with Cep-A2 (knot at [R.A., Dec]= [7$''$,18$''$] in Fig.~1). While these additional sources of gaseous CO$_2$ certainly contribute to the scatter seen in the right panel of Fig.~3, none of these sources can account for the correlation we observe between $N$(CO$_2$) and $N$(H$_2$) S(2) over such a large spatial extent. 

The low gas-phase CO$_2$ abundances detected by {\it ISO} ($\sim$10$^{-7}$ instead of the expected few$\times$10$^{-6}$) toward active star-forming regions with large CO$_2$ ices reservoirs (e.g. van Dishoeck et al. 1996) fostered theoretical efforts to investigate the effects that shocks might have on CO$_2$ chemistry. Charnley \& Kaufman (2000) showed that for shocks faster than a critical shock speed, which is a function of preshock gas density, CO$_2$ molecules sputtered off grain mantles are expected to be efficiently destroyed in the shock through reactions with H and H$_2$. On the other hand, a shock speed higher than 15 km s$^{-1}$ is necessary to allow efficient sputtering. 

Comparison of our results with the Charnley \& Kaufman (2000) shock-model predictions indicates that slow C-type shocks with speeds of at least $\sim$ 15 km s$^{-1}$ interacted with the ambient gas over the extent of the CO$_2$-emitting region (see Fig~1). Adopting a pre-shock density of $n_H \sim$ 10$^{5}$ cm$^{-3}$ (Goetz et al. 1998), our abundance limit further indicates that the shock speed is no greater than 30 km s$^{-1}$ since, at this density, higher shock speeds would efficiently remove CO$_2$ from the gas phase, mainly through reactions with hydrogen atoms (see Fig.~3 of Charnley \& Kaufman 2000). The CO$_2$ emission might arise at the flanks of the bow shock modeled by Froebrich et al. (2002). The average gas phase CO$_2$ abundance derived above (few $\times$10$^{-7}$) is consistent with a scenario in which a few percent of the grain material has been subject to shocks fast enough to sputter CO$_2$ ice mantles (with CO$_2$(ice)/H$_2$ $\sim$ few $\times$10$^{-5}$).





\acknowledgments

This work, which was supported in part by JPL contract 960803 to the Spitzer IRS Instrument Team and by RSA agreement 1263841, is based on observations made with the {\it Spitzer Space Telescope}, which is operated by the Jet Propulsion Laboratory, California Institute of Technology under a NASA contract. We are grateful to J.D. Green and K-H Kim for helping with the SMART software and standard IRS reductions. D.A.N. and P.S. acknowledge funding from LTSA program NAG 5-13114 to the Johns Hopkins University. J.P. acknowledges support from IRAC team contract SV474011. We are grateful to J.M. Torrelles for providing us with the NH$_3$ map. We thank the referee for fruitful comments.

\clearpage

\begin{figure}
\plottwo{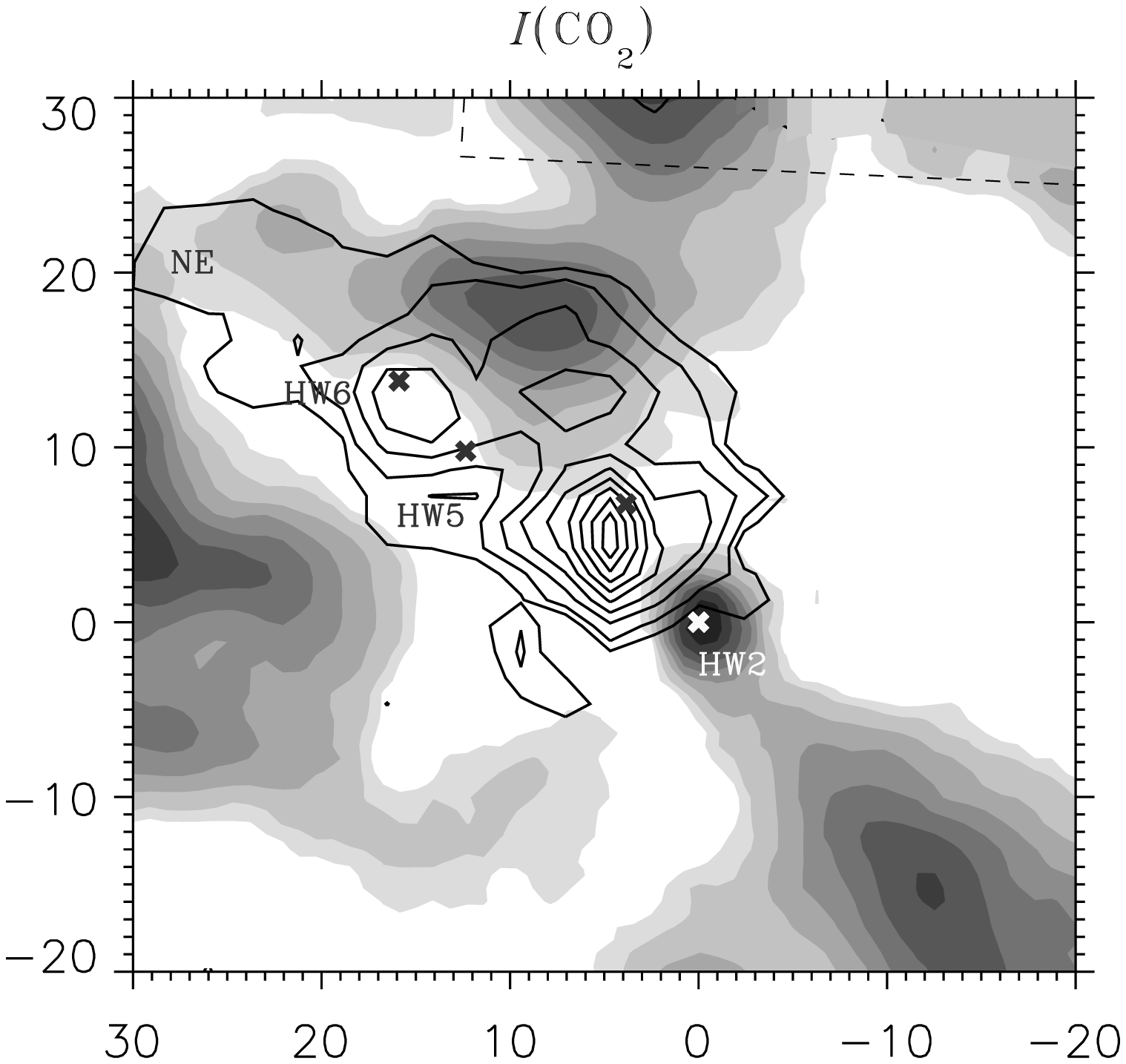}{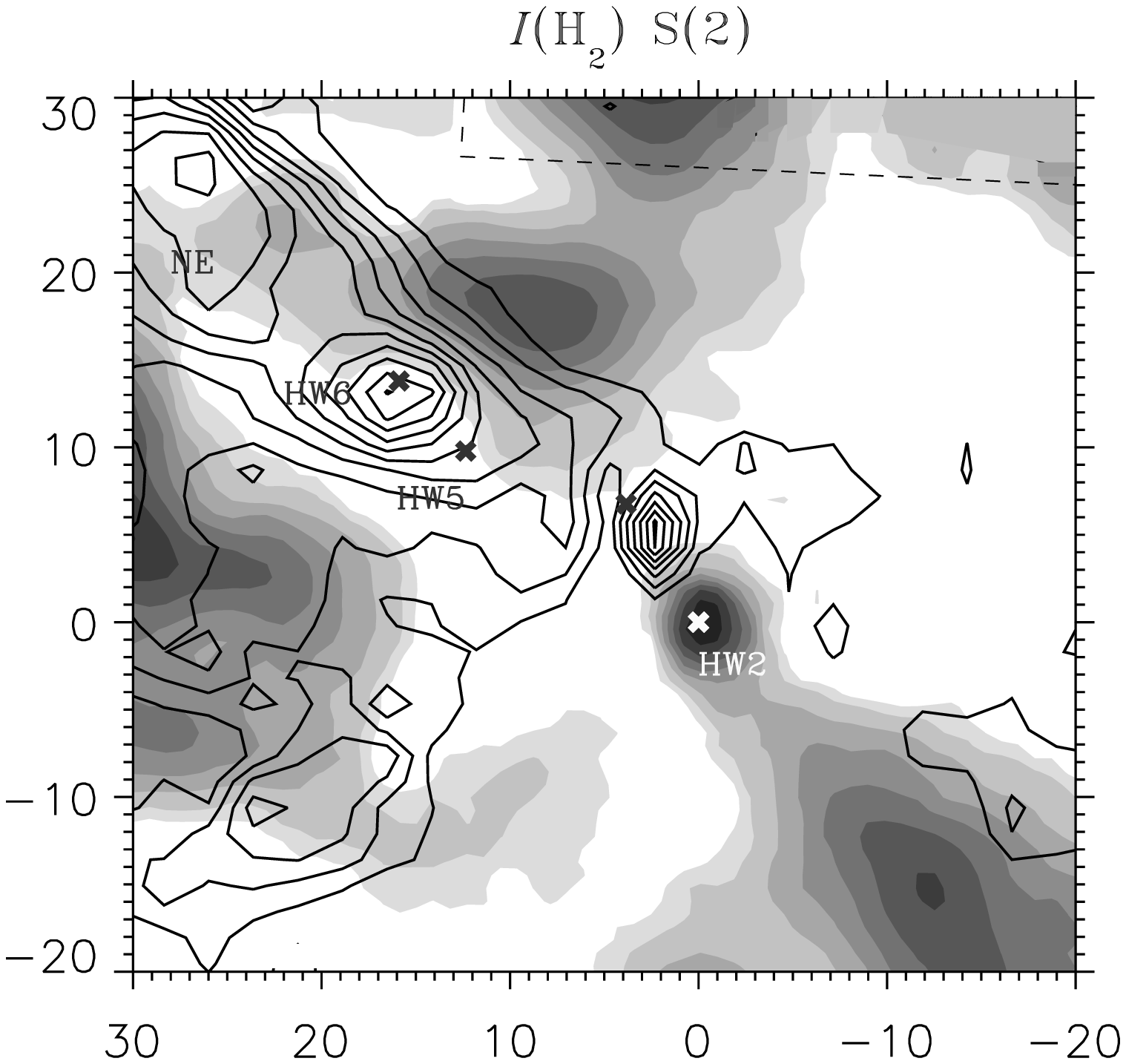}
\caption{Intensity distribution of gas-phase CO$_2$ emission ($Q$-branch at 14.98 $\mu$m) and H$_2$ S(2) pure rotational line emission (at 12.28 $\mu$m) toward Cepheus A East is shown as black contours. The lowest intensity contours are 0.75 and 1.0 for CO$_2$ and 1.0 for H$_2$ S(2) with steps of 0.5 $\times$ 10$^{-4}$ erg s$^{-1}$ cm$^{-2}$ sr$^{-1}$. The ($\times$) indicate the position of the radio-continuum sources HW2, HW4, HW5, HW6 and the EHV outflow edge (NE: $\Delta\alpha \cos\delta=$ 27$''$; $\Delta\delta=$ 22$''$). The coordinates are offsets in R.A. ($\Delta\alpha \cos\delta$) and declination ($\Delta\delta$) in arcsec with respect to HW2 (J2000: $\alpha=$22h56m17s.9 and $\delta= +$62$^{\circ}$01$'$49$''$; Hughes \& Wouterloot 1984). The grayscale shows the distribution of NH$_3$ (1,1), a tracer of the cold 
quiescent molecular gas. The lowest contours are 10 and 25 with steps of 25 mJy km s$^{-1}$ beam$^{-1}$ (Torrelles et al. 1993). The black dashed line delineates the mapped region in Short-High. \label{fig1}}
\end{figure}

\clearpage

\begin{figure}
\plottwo{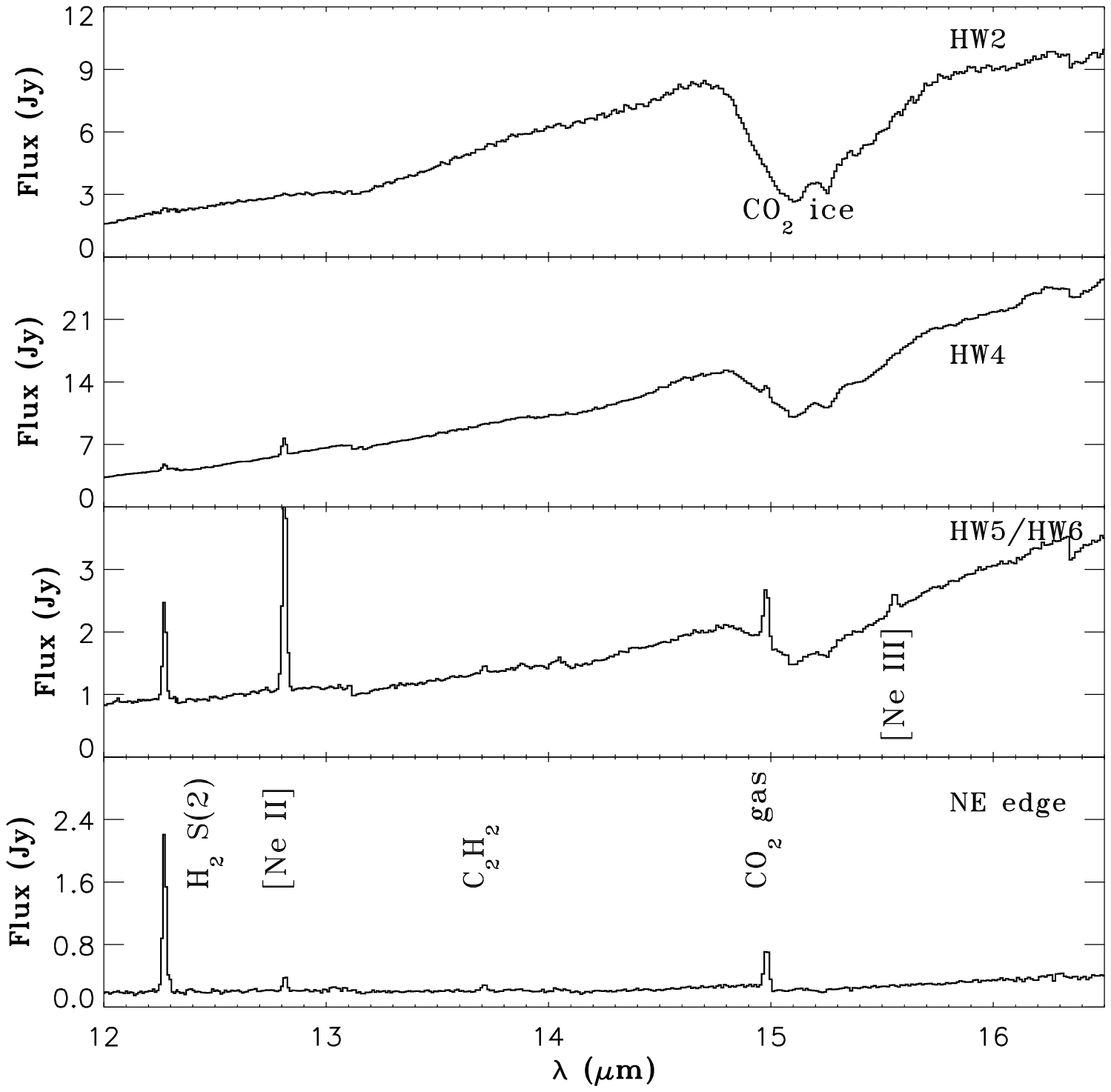}{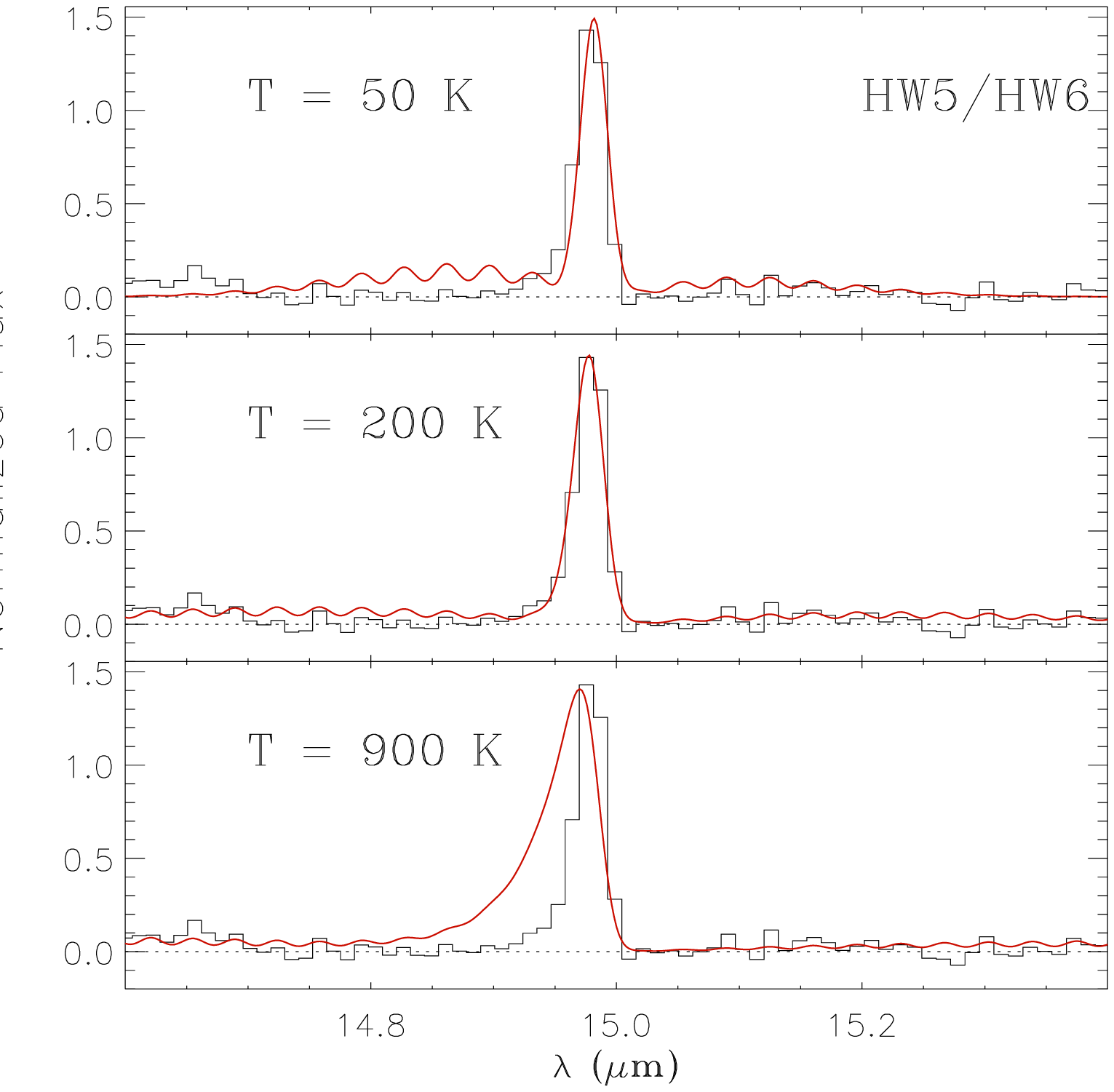}
\caption{ {\it left panel:} Summed IRS spectra of gas-phase CO$_2$, H$_2$, C$_2$H$_2$, [\ion{Ne}{2}] and [\ion{Ne}{3}] emission and CO$_2$ ice absorption toward Cepheus A East. Individual spectra were summed over regions $\sim$6$''$$\times$8$''$ in size, centered on the radio-continuum sources indicated in Fig.~1. Note the appearance of the gas-phase CO$_2$ emission feature and the progressive decrease in the CO$_2$ ice absorption feature when moving along the axis of the EHV outflow traced by the H$_2$ emission and the NH$_3$ cavities and away from HW2. Note also the presence of gaseous CO$_2$ emission alone at the NE position, the farthest from HW2 in our maps. {\it Right panel:} Continuum-subtracted summed spectrum from the HW5/HW6 position. Overimposed (red trace) is the calculated CO$_2$ band spectrum for $T=$ 50, 200 and 900K. Note the progressive shift of the bandhead and the $Q$-branch widening with increasing temperature. \label{fig2}}
\end{figure}

\clearpage

\begin{figure}
\plottwo{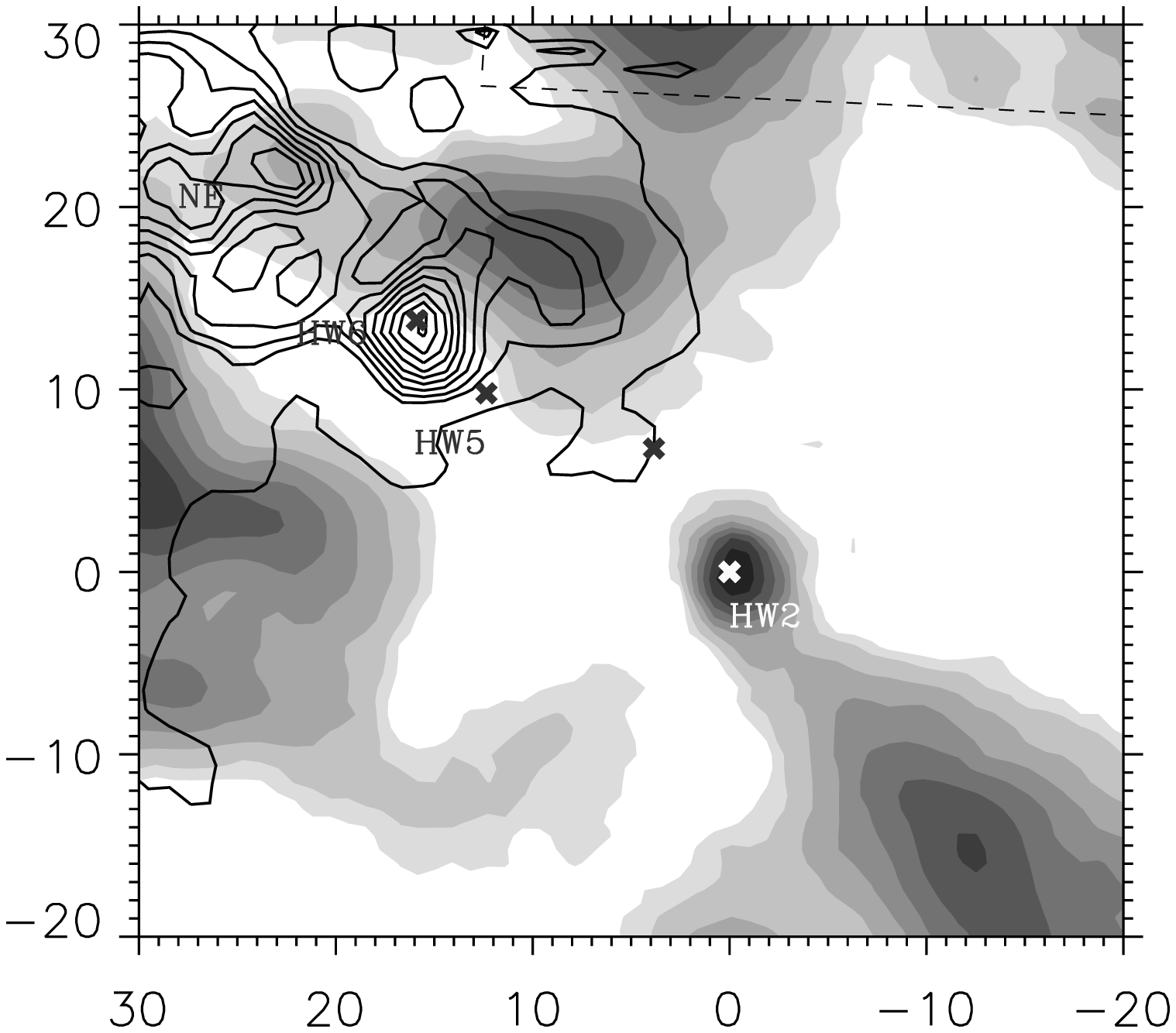}{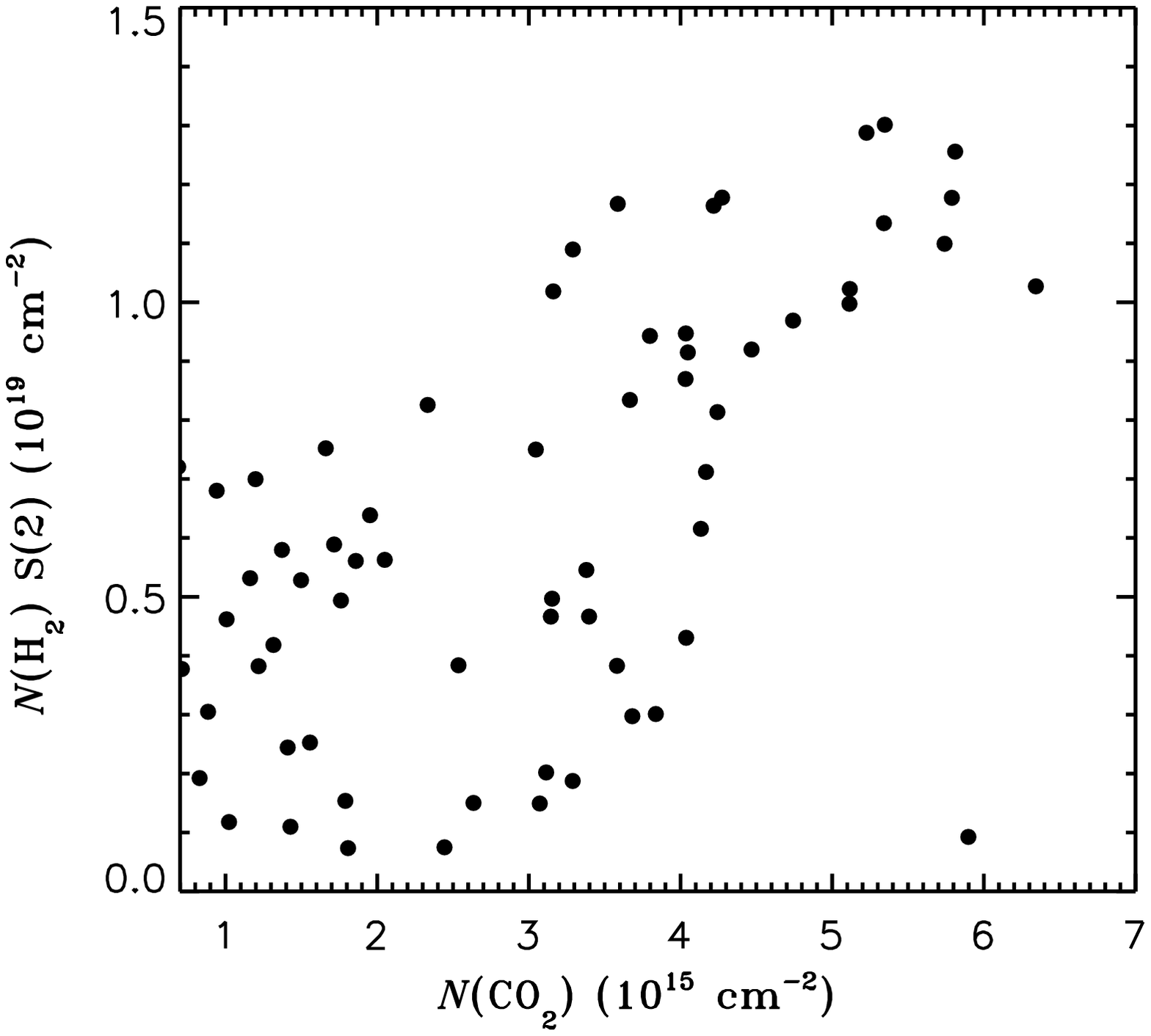}
\caption{{\it Left panel:} Distribution of gas-phase CO$_2$ column density, $N$(CO$_2$). The lowest contours are 1.5 and 3 with steps of 0.5$\times$10$^{15}$ cm$^{-2}$. The symbols and designations are as in Fig.~1. {\it Right panel:} $N$(CO$_2$) $vs$ $N$(H$_2$) $J=$ 4 (S(2) line) for the spatial positions exhibiting gaseous CO$_2$ ($Q$-branch) intensities starting at 0.75 $\times$10$^{-4}$ erg s$^{-1}$ cm$^{-2}$ sr$^{-1}$. \label{fig3}}
\end{figure}

\end{document}